\documentstyle[aps ]{revtex}
\begin{document}
\draft
\preprint{MKPH-T-98-1}
\title{Strange Isosinglet Versus Octet Scalar Axial Current }
\author{M. Kirchbach\\
{\small \it Institut f\"ur Kernphysik, J. Gutenberg Universit\"at, 
D-55099 Mainz, Germany} }
\maketitle
\begin{abstract}

The flavor SU(3)$_f$ group symmetry of QCD is systematically
considered as the heavy $c$ quark limit of SU(4)$_f$. Within that 
scheme we argue that the U(1)$_A$ anomaly
prevents Gell--Mann's choice for the realization of the su(4)$_f$
(and thereby the su(3)$_f$) 
algebra for the particular case of the neutral axial currents. 
Rather, Weyl's choice for the SU(4)$_f$ generators has to be used 
which leads to a neutral strong axial current having same structure
(up to a constant factor)  as the weak one. 


\noindent
KEY words: flavor symmetry, hypercharge axial current, U(1)$_A$ anomaly  

\end{abstract}

The SU(3)$_f$ flavor symmetry is one of the most important concepts
of contemporary hadron physics. It has its roots in the empirical observation 
that the lowest mass pseudoscalar mesons and spin-1/2 baryons 
join to octets, while the lightest spin--3/2 baryons and the 
lightest vector mesons constitute a decuplet, and a nonet,
respectively. From this point of view the existence of the first three 
fundamental flavor degrees of freedom of hadron matter, 
the $u $, $d $, and $s$ quarks, has been concluded. 
After the discovery of the charmed $(c)$ quark, the extension 
of SU(3)$_f$ to SU(4)$_f$ was considered despite of the apparent
mismatch between the mass of the $c$ quark on the one side, and the
masses of the $u,d$, and $s$ quarks, on the other side. Nonetheless,
this extension is useful because the SU(4)$_f$ group represents a 
symmetry operation acting onto two complete quark generations and 
thus has common representations with the electroweak symmetry. 
Henceforth, the SU(3)$_f$ group which still preserved its 
importance as the relevant flavor symmetry of QCD, 
will be considered to emerge in the heavy c--quark limit of SU(4)$_f$.

The algebra of the four--flavor special unitary group is determined by the 
fifteen generators $G^i$ satisfying the commutation relations
\begin{equation}
\lbrack G^i ,G^j\rbrack =
2if_k^{ij}\, G^ k \, ,\qquad i,j,k =1,..., 15\, .
\label{su4_alg}
\end{equation}
Note that the structure constants $f^{ij}_k$
$= tr(\lbrack G^i,G^j\rbrack G_k )/4i $ (they are completely antisymmetric 
in the indices $i,j,$ and $k$) depend on the explicit realization of the 
generators $G^i$ as linearly independent traceless Hermitian operators. 

The most natural realization of an su(n) Lie algebra is constructed 
within the (n$^2$-1)--dimensional basis 
defined in terms of the matrices $E_{ik}$ as \cite{Cornw}
\begin{eqnarray}
\langle m|E_{ik}| n \rangle &=& \delta_{in}\delta_{mk}\, ,
\qquad i,k =1,..., n\, ,\qquad i\not=k\, , \nonumber\\
E'_{tt} &=& E_{tt} -E_{t+1, \, t+1}\, , \qquad t=1,...,n-1\, , \nonumber\\
\lbrack E_{ik}, E_{lm}\rbrack &=& \delta_{im} E_{lk} -\delta_{lk} E_{im} \, .
\label{weyl_basis}
\end{eqnarray}
In other words, $E_{ik}$ is a matrix containing the unit at the 
intersection between the $k$th row and the $i$th column, while all the other 
matrix elements are equal to zero. For the diagonal 
matrices $E'_{tt}$ one finds $(E'_{tt})_{tt} =1$, and $(E'_{tt})_{t+1,t+1}= -1$, 
respectively. The so--called Weyl choice for the SU(4) generators is now
given by 
\begin{eqnarray}
G^r  =  E_{lm} + E_{ml} \, , \qquad r=1,...,6, && 
                                 \quad 
\mbox{for}
\quad l< m, m=2,..,4\, ,  
\nonumber\\
G^s = E'_{k,k } \, , \qquad s= 7,8,9,  &&
                                   \quad \mbox{for}\quad  k=1,2,3  \, , 
\nonumber\\
G^t = i(E_{lm}-E_{ml}) \, , \qquad t=10,...,15 \, , &&
                                 \quad \mbox{for} \quad  
 l< m \, , m=2,...,4\, .  
\label{gen_weyl}
\end{eqnarray}

Note that any set of 15 matrices obtained as linearly independent 
combinations of the Weyl matrices defined above can be considered as 
a realization of the su(4)$_f$ algebra.
{}For example, one can construct a new su(4)$_f$ algebra 
in keeping all the non--diagonal elements unchanged while
replacing the diagonal elements $E_{11}'$, $E_{22}'$ and $E_{33}'$
by the respective matrices $\Lambda^3$, $\Lambda^8$, and $\Lambda^{15}$ 
introduced as
\begin{eqnarray}
\Lambda^3=E_{11}' &=&
\left(\begin{array}{cccc}
1&\,\, \,\, 0&0&0\\
0&-  1&0&0\\
0&\,\,\, \, 0&0&0\\
0&\,\, \, \, 0&0&0
\end{array}\right)\, ,\nonumber\\
\Lambda^8 = {1\over \sqrt{3}}\, (E_{11}'+2E_{22}'+2E_{33}')\, =
{1\over \sqrt{3}} \left(\begin{array}{cccc}
1& 0&0&\,\,\,\, 0\\
0& 1&0&\,\, \, \, 0\\
0& 0&0&\, \, \, \, 0\\
0& 0&0&-2
\end{array}\right)\, ,
&\quad & \Lambda^{15} = {1\over \sqrt{6}}\, (E_{11}'+2E_{22}'-E_{33}')\, 
={1\over \sqrt{6}}
\left(\begin{array}{cccc}
1&0&\,\,\, \, 0&0\\
0&1&\, \, \, \, 0&0\\
0&0&  -3&0\\
0&0&\, \, \, \, 0&1
\end{array}\right)\, .\nonumber\\
\label{Lambdas_Es}
\end{eqnarray}
This new set of SU(4)$_f$ generators is known from standard 
textbooks as Gell-Mann's choice for the su(4)$_f$ algebra 
(see Ref.~\cite{Itz}). The question which one of these two algebras is 
relevant for hadron systems can be answered only by comparison with suitable
physical observables. We here show that, while both algebra choices appear 
equivalent for the vector currents, this is not the case for the axial vector currents.
In the latter case, the U(1)$_A$ anomaly prevents the existence of an
octet axial current and thereby Gell-Mann's choice for the
su(4)$_f$ algebra. Our argumentation goes as follows.

The main physical manifestations of unitary symmetries are twofold. 
These are the particle multiplets, on the one side, 
and the Noether currents, on the other side.
{}From the mathematical foundations of Lie groups it is well known,
that while the dimensionality of the irreducible representations of the 
group under consideration does not depend on the concrete realization
of its generators, the Noether currents do.
To gain an insight into the choice for the diagonal
su(4)$_f$ elements relevant for hadron processes one 
has, therefore, to analyze the flavor structure 
of the fundamental electromagnetic (vector) current, $J_\mu $ of the quarks.  
In doing so, the following decomposition of $J_\mu $ is found in the Weyl basis:  
\begin{eqnarray}
J_\mu & =&  {2\over 3}\bar u \gamma_\mu u 
               -{1\over 3} \bar d \gamma_\mu d
               +{2\over 3} \bar c \gamma_\mu c
               -{1\over 3} \bar s \gamma_\mu s \nonumber\\ 
&=& J_\mu^{11} + J_\mu^{33} + J_\mu^B\, , \nonumber\\
J_\mu^{11} := \bar q \gamma_\mu 
{  { E_{11}' } \over 2} q\, ,
\quad 
J_\mu^{33} &:=& \bar q \gamma_\mu {  { E_{33}'  }\over 2} q\, ,
\qquad  J_\mu^B:=  {1 \over 2} \bar q 
{{1\!\!1_4}\over 3} \gamma_\mu q\, ,
\qquad \bar q= (\bar u,\bar d,\bar c,\bar s ).
\label{elm_curr}
\end{eqnarray}       
Here, the current $J_\mu^{11}$ is precisely the 
thrid component of the ordinary isospin current,
while $J_\mu ^B$ is half the baryon number current.
Special attention has to be paid to 
the last current $J_\mu^{33}$ appearing diagonal
in the fundamental quark quadruplet.
This current is non--vanishing  only for the heavy flavor 
$c$ and $s$ quarks from the second quark generation, and its 
diagonal matrix elements in turn equal ${1\over 2}C$, and ${1\over 2}S$, with 
$C$ and $S$ standing in turn for the respective charm and strangeness 
quantum numbers of the quarks. 

{}From Eq.~(\ref{elm_curr}) follows that 
the fundamental vector quark current $J_\mu$ transforms according to the
sum of the singlet representation $\lbrace 0\rbrace $ with
the 1st and 2nd quark generation doublets
$\lbrace 2\rbrace _{ud}$, and $\lbrace 2\rbrace _{cs}$, respectively, rather
than according to the fundamental quark quadruplet $\lbrace 4\rbrace $, i.e. 
one finds $J_{\mu }$ to transform according to
\begin{equation}
\lbrace 4\rbrace \to \lbrace 2\rbrace_{ud} \oplus \lbrace 2\rbrace_{cs} 
\oplus \lbrace 0\rbrace  \, . 
\label{raspad}
\end{equation}
This decomposition shows that the symmetry of the QCD Lagrangian in
the neutral sector is in fact not the full SU(4)$_f$ group but rather 
its restriction to SU(2)$_{ud}\otimes $SU(2)$_{cs} \otimes $U(1)$_V$.
Here the SU(2)$_{ud}$ and SU(2)$_{cs}$  groups act in turn onto the 
1st and the 2nd quark generations. On the level of the algebra one has 
\begin{equation}
su(4)_f\to su(2)_{ud}\oplus su(2)_{cs}\oplus u(1)_V\, .
\label{sushenie}
\end{equation}
As long as both $J_\mu^{33}$ and the baryon current are conserved,
one can introduce their sum, here denoted by $J_\mu^Y$,
as a new conserved current 
\begin{equation}
J_\mu^Y = J_\mu^{33}+J_\mu^B\, .
\label{su2cd_u1}
\end{equation} 
The total flavor charge, $Q^Y(t) $, of that current  
\begin{equation}
Q^Y (t) = \int_0^\infty J_0^Y(t, \vec{x}\, )\mbox{d}^3\vec{x}\, ,
\label{octet_current}
\end{equation}
will be a constant of motion and can be simultaneously diagonalized with
the Hamiltonian of the hadron system.
Its eigenvalues for the 1st and 2nd quark generations are given in turn
by the following expressions,
\begin{equation}
Q^Y\,\left(\begin{array}{c}
c\\
s\end{array}\right) = 
\left(\begin{array}{cc}
C+ {1\over 3}&0\\
0&S+{1\over 3}
\end{array}\right)\,
\left(\begin{array}{c}
c\\
s\end{array}\right)\, ,
\qquad
Q^Y\,\left(\begin{array}{c}
u\\
d\end{array}\right) = 
\left(\begin{array}{cc}
{1\over 3}&0\\
0&{1\over 3}
\end{array}\right)\,
\left(\begin{array}{c}
u\\
d\end{array}\right)\, ,
\label{huper_charge}
\end{equation}
and can be used to label the hadron states.
{}From Eq.~(\ref{elm_curr}) follows that the total vector charge operator 
$\hat{Q}_V(t)=\int_0^\infty J_0(t,\vec{x}\, )
\mbox{d}^3\vec{x}$ is related to the operators of isospin 
$\hat{t}_3 ={1\over 2} E_{11}' $, and hypercharge,
$\hat{Y} = E_{33}'+{1\over 3}1\!\!1_4 $, via
\begin{eqnarray}
\hat{Q}_V = \hat{t}_3 +{\hat{Y}\over 2}
&=&{1\over 2} \left(\begin{array}{cccc}
1+{1\over 3}&0&0&0\\
0&-1 +{1\over 3}&0&0\\
0&0&C+{1\over 3}&0\\
0&0&0&S+{1\over 3}\, ,
\end{array}\right)
\label{GMn_Nishi}
\end{eqnarray}
thus giving rise to the Gell-Mann--Nishijima relation.
Eq.~(\ref{GMn_Nishi}) clearly illustrates that
four--flavor hypercharge can not be introduced on the SU(4)$_f$
level but rather requires the full U(4)$_f$ group.

Now in freezing the charm degree of freedom
of the hypercharge current in Eq.~(\ref{su2cd_u1}),  Gell-Mann's 
octet vector current is reproduced in the truncated flavor space as
\begin{equation}
{1\over {2\sqrt{3}}}J_\mu^8 = \lim_{m_c\to \Lambda_c} \, J_\mu^Y\, ,
\qquad \Lambda_c \approx 1.5\,\, \mbox{GeV} \, .
\label{GellManns_choice}
\end{equation} 
Here $\Lambda_c $ has been chosen to be sufficiently large in order
to ensure negligible $c$ quark effects on the 1 GeV mass scale, on the
one side, but finite, in order to preserve the anomaly free character of 
the SU(4)$_f$ theory, on the other side \cite{Aoern}.
Eq.~(\ref{GellManns_choice}) creates the impression, that for the 
particular case of three flavors, hypercharge may be introduced within 
the special group. In fact this is not correct as the essential
ingredient of the hypercharge is the baryon number 
which, in being associated with the conservation of the U(1)$_V$ current,
can not be defined within the SU(3)$_f$ framework.
In other words, if $\Lambda ^8$ is to enter the SU(3)$_f$ version of the
Gell-Mann-Nishijima relation and be interpreted as the hypercharge generator, 
it has to be defined instead by Eq.~(\ref{Lambdas_Es}) rather by
\begin{equation}
{1\over \sqrt{3}}\, \Lambda^8 
=\lim_{m_c \to \Lambda_c} \left( E_{33}' +{1\over 3}\, 
1\!\!1_4\right)\,  .
\label{u4_lambda8}
\end{equation}
Note that by means of $\Lambda^8$ from Eq.~(\ref{Lambdas_Es}) 
the electric charges of the quarks would be obtained as
an artificial combination of third projections of isospin ($t_3$) an U-spin 
($t_3^U:=E_{22}'/2$) quantum numbers according to 
$Q=t_3 + {1\over 3}(t_3 + 2t_3^U)$ rather than by the genuine 
Gell-Mann--Nishijima relation in Eq.~(\ref{GMn_Nishi}).
Until one is concerned with the anomaly free vector currents, however,
Gell--Mann's matrix $\Lambda^8 $ can be viewed as an 
{\it effective\/} representation of hypercharge within the 
truncated flavor space. In such a case, using Gell-Mann's choice for the 
su(3)$_f$ algebra doesn't create any problems and is equivalent to Weyl's 
choice. In the axial sector, however, Gell-Mann's choice 
is of minor use as the axial transformation associated with
the unit matrix from Eq.~(\ref{u4_lambda8}) will transport the
problems of the U(1)$_A$ anomaly onto the octet axial current.

The observation that the octet matrix $\Lambda^8$ has no
independent fundamental meaning as an element of the su(4)$_f$
algebra is further strongly supported by the 
internal flavor structure of the
neutral vector mesons which are evidently Weyl bosons
as their states transform
in accordance to SU(2)$_{ud}\otimes$ SU(2)$_{cs}\otimes$ U(1)$_V$
representations:  
\begin{eqnarray}
|\rho^0\rangle = {1\over \sqrt{2} } (\bar u\,\,  \bar d )E_{11}'
\left( 
\begin{array}{c}
u\\
d\end{array}\right)
\, , &\quad &
|\omega \rangle  = {1\over \sqrt{2} } 
(\bar u\,\,  \bar d ) 1\!\!1_2 
\left(\begin{array}{c}
u\\
d\end{array}\right)\, , \nonumber\\
|\phi\rangle  &= & \lim_{m_c \to \Lambda_c }\, \,
(\bar c \,\, \bar s)\, E_{33}'\, 
\left(\begin{array}{c}
c\\
s\end{array}\right)\, .
\label{vector_mesons}
\end{eqnarray}
The well known 'ideal mixing' within the vector meson nonet therefore
corresponds to the rotation of Gell-Mann's to Weyl's basis.
Note that the separation of the strange and non--strange quarkonia within the 
vector meson nonet is known as the Okubo-Zweig-Iizuka (OZI) rule and reflects 
the existence of gluon exchange between the quarks.
The OZI rule is therefore an expression for favoring Weyl's 
over Gell-Mann's choice for su(4)$_f$.

Now one has to answer the question whether or not Gell-Mann's choice
{}for the elements of the su(3)$_f$ algebra will apply to the axial vector
currents. To answer this question let us first recall that
the generators of axial flavor transformation
don't constitute an algebra by themselves. They can mostly be considered as 
ingredients of group transformations when associating them to the
vector flavor transformation to produce the respective 
left (L) and  right (R) chiral rotations. For this reason, the axial currents
{\it have to replicate in flavor space the structure of the vector currents\/}.
In other words, the neutral chiral hadron currents will also transform
according to the representation of 
\begin{equation}
\lbrack  su(2)_{ud}\, ^{(L)} \oplus su(2)_{cs}\, ^{(L)}\rbrack
\oplus
\lbrack su(2)_{ud}\, ^{(R)} \oplus su(2)_{cs}\, ^{(R)}\rbrack
\oplus
\lbrack u(1)_V\oplus u(1)_A\rbrack \, .
\label{chiral_alg} 
\end{equation}
Within the context of Eq.~(\ref{chiral_alg}),
the only neutral axial current having a well defined chiral limit will be  
\begin{equation}
J_{\mu ,5} = \bar q \gamma_\mu\gamma_5 {{E_{11}'}\over 2}q
             +\bar q \gamma_\mu\gamma_5 {{E_{33}'}\over 2}q\, .
\label{weak_curr}
\end{equation}
The decomposition of $J_{\mu , 5}$ in the last equation  
reflects the exclusion of the anomalous U(1)$_A$ current \cite{anom} 
which can not be used any longer as a building block for the construction 
of an anomaly free octet axial current.

One remarkable feature of $J_{\mu ,5}$ is that its structure 
is identical (up to the factor of $-1/2$) to that of the weak 
axial vector current. For this reason,
the well established universality of the charged weak and strong axial 
vector currents underlying the current algebra can be extended to include 
the neutral ones. In the heavy $c$ quark limit, the  $J_{\mu ,5}$  current 
decomposes into an isovector ($J_{\mu , 5}^I$) and a purely strange 
SU(2)$_I$ isosinglet ($J_{\mu , 5}^s $) component 
\begin{eqnarray}
J^I_{\mu ,5} = \bar q \gamma_\mu \gamma_5 {{E_{11}'}\over 2} q \, ,
&\qquad &
 J^s_{\mu ,5} = \lim_{m_c\to \Lambda_c}  
(\bar q  \gamma_\mu\gamma_5 {{E_{33}'}\over 2} q )\, . 
\label{isov_isosc}
\end{eqnarray}

In contrast to the physical vector mesons $\phi $ and $\omega $,
their corresponding $\eta $ and $\eta '$ mesons from the
pseudoscalar nonet are not Weyl bosons respecting the OZI rule.
Within the $0^-$ nonet the U(1)$_A$ anomaly leads to
a significant mixing of the Weyl states and thereby to
a violation of the OZI rule \cite{Shur}. 
{}For the particular case of quadratic mass formulae,
the physical $\eta $ meson turns out to emerge in the
heavy $c$ quark limit from the following Weyl states mixing: 
\begin{equation}
|\eta \rangle= \lim _{m_c\to \Lambda_c}\,\,
\cos \epsilon  \, (\bar c c -\bar s s)
-\sin\epsilon \, {1\over \sqrt{2} } (\bar u u + \bar d d)\, ,
\label{phys_eta}
\end{equation} 
with $\epsilon \approx -45,4^0$. This is an alternative way to
express the tendency of the flavor structure of the $\eta $ meson 
towards Gell-Mann-Okubo mass formulae predictions. 
In other words, while the wave function of the $\eta $ meson 
can be roughly associated with the scalar of the octet,
the $\eta N N$ vertex will be an isosinglet rather than an F-spin scalar 
because the $\eta $ meson has to couple to the isosinglet nucleon current,
\begin{equation}
J_{\mu ,5}^{s (N)} = \langle N| {1\over 2}\bar s \gamma_\mu\gamma_5 s
|N\rangle = G_1^s \bar {\cal U}_N\gamma_\mu\gamma_5{{1\!\!1}\over 2} 
{\cal U}_N\, ,
\qquad G_1^s = \Delta s\, ,
\label{G1_s}
\end{equation}
which can happen only via its strange quarkonium component. In assuming, 
{}for simplicity, the wave function of the $\eta $ meson to be the pure
scalar of the octet, its isosinglet axial current is now defined as
\begin{eqnarray}
J_{\mu , 5}^{s\,  (\eta )} =\langle 0|{1\over 2}\bar u 
           \gamma_\mu\gamma_5 u
         -{1\over 2} \bar d \gamma_\mu\gamma_5 d 
         -{1\over 2} \bar s \gamma_\mu\gamma_5 s 
| {1\over \sqrt{6} } (\bar u u + \bar d d - 2\bar s s ) \rangle 
&=& f_\eta\,  i q_\mu\, , \nonumber\\
f_\eta &:=& {1\over \sqrt{6} }\kappa_s (0^-)\,  m_\eta  \, .
\label{eta_axcurr}
\end{eqnarray}
Here $\kappa_s (0^-)$ denotes the dimensionless coupling of the strange
quarkonium component of the pseudoscalar meson to the strange axial 
vector current, while $f_\eta $, $m_\eta $ and $q_\mu $ are in turn the 
weak decay coupling, the mass and the momentum of the $\eta $ meson. 
{}Furthermore, the couplings between $\bar u u$ and $\bar d d $
quarkonia and quark axial currents have been suggested to be diagonal in 
flavor and universal in strength,
i.e. $\kappa_u (0^-) = \kappa_d (0^-)$, to recover isospin symmetry
already at the tree level\footnote{ The empirically 
observed closeness of the $\pi $ and $\eta $ weak decay constants,
($f_\eta \approx 1.1 f_\pi $) does not
necessarily imply $\kappa_u (0^-) \approx \kappa_s (0^-)$ at the tree 
level.}. With this in mind, the contact  $\eta NN $ vertex takes the form
\begin{eqnarray}
{\cal V}_{\eta N N } &= & {1\over f^2_\eta } \, 
                       J^{s\, (N)}\cdot J^{s\, (\eta )}\,  \nonumber\\
&=& {1\over f_\eta^2 } 
\left(G_1^s\bar {\cal U}_N\gamma\gamma_5 
{{1\!\!1}\over 2}{\cal U}_N \,\right) \cdot \, f_\eta \, iq \phi_\eta\, .
\label{vert_etaNN}
\end{eqnarray}
Here, the parameter of dimensionality $\lbrack mass^2 \rbrack $ entering 
the current--current coupling has been chosen to equal the weak $\eta $ 
decay coupling constant. This choice is convenient because of the 
universality of the isosinglet axial current and the related
ansatz on the meson pole dominance. It allows one to express the coupling 
of neutral pseudoscalar mesons to arbitrary targets in terms of their 
weak decay couplings to the vacuum.
The combination ${G_1^s\over {2f_\eta }}$ is conventionally denoted
by ${f_{\eta NN}\over m_\eta} $ with $f_{\eta NN}$ being called the
gradient $\eta N$ coupling constant. In inserting Eqs.~(\ref{G1_s}) and
(\ref{eta_axcurr}) in the last expressions, one is led to
\begin{eqnarray}
{f_{\eta NN} \over m_\eta } &=&{ G_1^s\over {2f_\eta } }
= \sqrt{6} {{\Delta s}\over {2\kappa _s (0^-) m_\eta } } \, , \nonumber\\
f_{\eta NN} & = & \sqrt{ {3\over 2}} {{\Delta s}\over {\kappa_s (0^-)}}\, .
\label{etaN_vert}
\end{eqnarray}     
This means that at tree level the contact gradient $\eta N$ coupling 
appears proportional to the fraction of proton helicity carried by the 
strange quark sea, rather than to the octet axial vector coupling 
$g_A^{(8)} ={1\over \sqrt{3}}(\Delta u + \Delta d - 2\Delta s)$, 
a result already conjectured in a previous work \cite{KiWe}. This might be 
one of the main reasons for which a strong suppression of the gradient 
$\eta N$ coupling has constantly been found over the years by various data 
analyses regarding the $\eta $ photoproduction off proton at threshold 
\cite{TiKa}, the $\bar p p $ collisions \cite{Kroll}, as well as the 
nucleon-nucleon (NN) and nucleon--hyperon (NY) phase shifts \cite{Reuber}. 

The essence of the present study is that to determine the $\eta N$ coupling
one has to exploit the neutral current in Eq.~(\ref{weak_curr}) which
respects the OZI rule and can be considered formally to emerge from
an ideal mixing between Gell-Mann's octet and flavor singlet
axial currents.
That $\eta_8 -\eta _0$ mixing will lead to a reduction
of the nucleon matrix element of the octet axial current was earlier
considered, for example, in Ref.~\cite{Ven}. There, however,
the reduction reported was not as big as the one in Eq.~(\ref{etaN_vert}). 

In view of the small $\eta N$ contact coupling, loop corrections acquire 
importance. The effect of one--loop corrections on the suppression of
the nucleon matrix element of the octet axial vector
current has been considered in Ref.~\cite{Sav} within the framework of 
chiral perturbation theory. There, the strong dependence of the result 
on the nucleon --$\Delta $ mass splitting was revealed and the 
necessity for higher order corrections discussed. 
In Ref.~\cite{KiWe} vertex renormalization effects were considered 
in terms of nucleon and meson degrees of freedom rather than on the 
quark level. There, the significance of triangular corrections of the type 
$a_0 (980) \pi N$ for the enhancement of the $(\bar s s) N N$ vertex was 
revealed. These provide the coupling of the strange quarkonium component of 
the $\eta $ meson to the non--strange meson cloud surrounding the nucleon via the 
violation of the OZI rule and lead to an effective $\eta N$ coupling, less than 
half the value predicted by the constituent quark model, in line with data.

To conclude, it was shown that the only isoscalar axial vector
current having a really anomaly free chiral limit is 
$J_{\mu , 5}$ in Eq.~(\ref{weak_curr}) which,
in being a Weyl current, automatically respects the OZI rule.
Unfortunately, due to U(1)$_A$ anomaly effects,
the internal flavor structure of the $\eta $ meson is not in accord with 
$J_{\mu ,5}^s$. For these reasons, the physical $\eta $ meson can mostly
be considered as an approximate 'strange' Goldstone boson.
As the U(1)$_A$ anomaly was shown to prevent the construction
of an octet axial current with a well defined chiral limit, 
the interpretation of the $\eta $ meson as the octet Goldstone boson has to 
be dropped despite the presence of $(\bar u u +\bar d d$) components
in its wave function.

Helpful comments by Martin Reuter and Andreas Wirzba are kindly 
appreciated. 
Work supported by the Deutsche Forschungsgemeinschaft (SFB 201).

\end{document}